\documentclass[showpacs,preprintnumbers,amsmath,amssymb]{revtex4}
\usepackage[dvips]{graphicx}
\usepackage{dcolumn}

\begin{document}

\title{Josephson vortex lattices with modulation perpendicular to an in-plane magnetic field in layered superconductor}

\author{Ryusuke Ikeda and Hidehumi Nawata}

\affiliation{%
Department of Physics, Kyoto University, Kyoto 606-8502, Japan
}

\date{\today}

\begin{abstract}
In quasi low dimensional superconductors under {\it parallel} magnetic fields applied along a conducting direction, vortex lattices with a modulation of Fulde-Ferrell-Larkin-Ovchinnikov (FFLO) type {\it perpendicular} to the field may occur due to an enhanced paramagnetic depairing. As the strength of an in-plane field is varied in a Q2D material, the Josephson vortex lattices accompanied by nodal planes are formed in higher Landau level (LL) modes of the superconducting (SC) order parameter and show field-induced structural transitions. A change of orientation of nodal planes induced by these transitions should be observed in transport measurements for an out-of-plane current in real superconductors with point disorder effective on the SC layers. Further, the $H_{c2}$-transition from this higher LL state to the normal phase is of second order for moderately strong paramagnetic effects but, in the case with a strong enough paramagnetic effect, becomes discontinuous as well as the transition between this modulated state and an ordinary Abrikosov vortex lattice in the lowest LL. Relevance of these results to recent observations in organic superconductors suggesting the presence of an FFLO state are discussed. 
\end{abstract}


\maketitle


Recent experimental evidences of a new type of high field phase in a heavy fermion superconductor CeCoIn$_5$ have led to a new occasion on studies of the so-called Fulde-Ferrell-Larkin-Ovchinnikov (FFLO) superconducting (SC) state \cite{Bianchi}. In CeCoIn$_5$ in ${\bf H} \parallel ab$, the features of phase diagram and a change of elastic response through the Abrikosov to FFLO transition were consistent with the picture based on the presence of an FFLO modulation parallel to the applied field ${\bf H}$ (and nodal planes perpendicular to ${\bf H}$) \cite{AI,IAPRL}. Reflecting the fact that the Fermi surface relevant to the superconductivity is not purely cylindrical in CeCoIn$_5$, the corresponding state has also been observed in ${\bf H}$ perpendicular to the layers (${\bf H} \parallel c$). \cite{Kumagai} 
This FFLO vortex state modulating along ${\bf H}$ has no additional variation of the vortex structure in the plane perpendicular to ${\bf H}$ and is, just like the ordinary Abrikosov lattice, well described in the lowest Landau level ($n=0$ LL) where no spatial variation other than the field-induced vortices occurs as far as the SC order parameter is a single scalar field. \cite{RI1,RI2} Actually, any vortex state stable in the $n=0$ LL is isotropic in character in the plane perpendicular to ${\bf H}$. 

Inevitably, possible FFLO states modulating in a direction {\it perpendicular} to the field have to be described in terms of higher LL modes of the SC order parameter. When imagining an artifitial situation in which the field strength inducing the formation of vortices is much weaker than the total magnetic field associated with the Zeeman energy, it is a {\it higher} LL state which determines the $H_{c2}(T)$-line and describes the vortex state just below it \cite{Klein}. It has been recently noticed that even the 3D high field vortex lattice may become such a higher LL state in clean limit and at low enough temperatures \cite{RI1}. Although such a state is easily lost if a finite quasiparticle damping is not negligible \cite{RI1}, it may be realized in superconductors with strong enough paramagnetic depairing. Among them, the $n=1$ LL state closer to the $n=0$ LL is the most relevant to real systems. It has a one-dimensional stripe pattern of nodal planes appearing separately from the vortices \cite{Klein,Macd} and is expected to have a peculiar property that directional probes such as the transport measurements become anisotropic depending on the orientations of the nodal planes relative to the crystal lattice. When considering the striped vortex lattices in layered systems in fields parallel to the layers, a question arises: A typical Fermi surface of such materials, i.e., the cylindrical one, 
suggests that the FFLO modulation tends to become parallel to the layers, accompanied by nodal planes vertical to the layers, while the nodal planes tend to be pinned, as well as the vortices, by the layer structure so that the layered structure favors nodal planes oriented along the layers. In considering a possible FFLO-like state in quasi 2D materials, it is necessary to correctly resolve such a competition in the order parameter structure. 

In this work, we have studied the $n=1$ LL Josephson vortex lattices occurring in the layered system in parallel fields in the situations where, due to a strong paramagnetic depairing, this state is realized below $H_{c2}(T)$. We find that, as the field is varied, the orientation of nodal planes changes accompanying structural transitions of the vortex lattice itself. Since, in real materials, the pinning effect due to the point disorder on the SC layers is effective especially for nodal planes not parallel but perpendicular to the layers, such changes of orientation of nodal planes should affect the resistivity for currents perpendicular to the layers. Relevance of these results to the observation \cite{Uji} in the quasi 2D organic field-induced superconductor $\lambda$-(BETS)$_2$FeCl$_4$ will be discussed. In addition, possible phase diagrams in cases with the $n=1$ LL state in the parallel fields will also be discussed, and we point out that the $H_{c2}$-transition (i.e., the mean field SC transition) is of second order for reasonable values of the Maki parameter, while it becomes of first order for very high but, nevertheless, realistic values of the Maki parameter. This result may be relevant to the recent report of heat capacity data of a $\kappa$-(ET)$_2$ organic material \cite{Lortz}. 
 
We start from the same BCS model as in Ref.\cite{RI2} for quasi 2D systems which includes the Zeeman energy and the interlayer hopping energy terms
\begin{eqnarray}
\Delta {\cal H} &=& - d \sum_{\sigma, j}  \int d^2 r_{\perp} \Bigg[ \sigma \mu_B H (\varphi^\sigma_j({\bf r}_\perp))^\dagger \varphi^\sigma_j({\bf r}_\perp) + \frac{J}{2} \Bigg( 
({\varphi}_{j}^\sigma ({\bf r_\perp}))^\dagger 
{\varphi}_{j+1}^{\sigma}({\bf r_\perp}) +  ({\varphi}_{j+1}^\sigma({\bf r_\perp}))^\dagger {\varphi}_{j}^{\sigma}({\bf r_\perp}) \Bigg) \Bigg],
\label{interlayer}
\end{eqnarray}
where $j$ is the index numbering the SC layers, $d$ is the interlayer spacing, and $\sigma \mu_B H = \mu_B H$ or $-\mu_B H$ is the Zeeman energy. In discussing our calculation results, the strength of the paramagnetic effect will be measured by the Maki parameter, i.e., the ratio between the orbital and Pauli limiting fields, which will be defined here by the quantity $\alpha_M = \mu_B H^{({\rm orb})}_{\rm 2D}(0)/k_B T_c$. The conventional Maki parameter in ${\bf H} \parallel c$ 
is obtained by multiplying a constant factor $\sim 7.0$ to this $\alpha_M$, 
where $H^{({\rm orb})}_{\rm 2D}(0) \sim 1/(2e \xi_0^2)$ is the orbital limiting field in 2D limit, and $\xi_0$ is the in-plane coherence length. Hereafter, the applied field ${\bf H}$ is directed to the $x$-axis parallel to the SC layer. 

In studying nearly 3D-like superconductors in which the out-of-plane coherence length $\xi_c(0)$, which will be defined later, is longer than $d/\sqrt{2}$, the interlayer hopping energy term $\propto J$ is treated on the same footing as the in-plane kinetic energy term, and, instead, effects of the discrete layered structure on the SC order parameter are not well incorporated in the GL description \cite{IAPRL,RI1,RI2}. Since this layering effect on the SC order parameter is one of the main concerns in this work, we choose here rather to treat $J$ perturbatively. When the SC order parameter belongs primarily to the $n$-th LL, the resulting GL free energy in the mean field approximation takes the form 
\begin{eqnarray}\label{eq:LD}
{\cal F}_{\rm LD}^{(n)} &=& d \sum_j \int dy \biggl[ (\Delta_n^{(j)}(y))^* a_n \Delta_n^{(j)}(y) 
+ (\Delta_n^{(j)}(y) - \Delta_n^{(j+1)}(y))^* c_n (\Delta_n^{(j)}(y) - \Delta_n^{(j+1)}(y))  \nonumber \\ 
&+& \frac{V_{4,n}({\bf \Pi}_s)}{2} (\Delta_n^{(j)}(y_1) \, 
\Delta_n^{(j)}(y_3))^* \Delta_n^{(j)}(y_2) \, \Delta_n^{(j)}(y_4) 
|_{y_s \to y} 
\biggr]
\label{LD1}
\end{eqnarray} 
similar to the familiar Lawrence-Doniach model, where $\Delta_n^{(j)}$ implies the projection of $\Delta^{(j)}$ into the $n$-th LL, and ${\bf \Pi}_s = -{\rm i}\partial/\partial{\bf r}_s + 2e {\bf A}({\bf r}_s)$. If the $H_{c2}$-transition is discontinuous, O($|\Delta_n^{(j)}|^6$) term omitted in eq.(\ref{LD1}) needs to be incorporated. Microscopic details in the vortex states are reflected altogether in the coefficients such as $a_n$ and $c_n$. In eq.(\ref{eq:LD}), a possible modulation parallel to ${\bf H}$ in $\Delta_n$ was neglected. Possibilities of this modulation need to be incorporated in considering phase diagrams and will be discussed at the end of this paper. The LL representation of the order parameter can be used for the layered system by rewriting \cite{II,RI02} ${\cal F}_{\rm LD}^{(n)}$ in the form 
\begin{eqnarray}
{\cal F}_{\rm LD}^{(n)} &=& \int dz \int dy \sum_m \exp({\rm i}2 \pi m z/d) \biggl[ (\Delta_n(y,z))^* a_n \Delta_n(y,z) + (\Delta_n(y,z) - \Delta_n(y,z+d))^* c_n (\Delta_n(y,z) - \Delta_n(y,z+d)) \nonumber \\
&+& \frac{V_{4,n}({\bf \Pi}_s)}{2} (\Delta_n(y_1,z) \, 
\Delta_n(y_3,z))^* \Delta_n(y_2,z) \, \Delta_n(y_4,z) \biggr]. 
\end{eqnarray}
That is, spatial variations of $\Delta_n^{(j)}$ on the SC layers are described in terms of the continuous order parameter $\Delta_n(y,z)$. Hereafter, the linear gauge ${\bf A}= - H z {\hat y}$ will be used. 

The coefficients will be treated in the same manner as in Ref.\cite{RI2}. The coefficients $a_n$ and $c_n$ of the quadratic term are given by 
\begin{eqnarray}
a_n &=& \frac{1}{2}{\rm ln}(h) + \int_0^\infty d\rho \biggl[ \frac{1}{\rho} \exp\biggl(-\frac{\pi^2 \xi_0^2 \rho^2}{r_H^2} \biggr) - f(\rho) \biggl\langle |{\hat \Delta}_p|^2 L_n(|\mu|^2 \rho^2) \, \exp\biggl(-\frac{|\mu|^2 \rho^2}{2} \biggr) \biggr\rangle_{\rm FS} \biggr], 
\end{eqnarray}
where 
\begin{equation}
\mu=\frac{w_y 
+ {\rm i} w_z}{\sqrt{2} \, \, r_H T_c},
\label{mu}
\end{equation}
\begin{equation}
f(\rho) = \frac{2 \pi \, t}{{\rm sinh}(2 \pi t \rho)} \, {\rm cos}\biggl(\frac{2 \mu_B H\, \rho}{T_c} \biggr), 
\label{f}
\end{equation}
\begin{equation}
c_n = \biggl(\frac{J}{2 T_c} \biggr)^2 \int_0 d\rho \rho^2 f(\rho) \biggl\langle |{\hat \Delta}_p|^2 L_n(|\mu|^2 \rho^2) \, \exp\biggl(-\frac{|\mu|^2 \rho^2}{2} \biggr) \biggr\rangle,
\label{cn}
\end{equation}
and ${\hat \Delta}_p$ denotes the normalized orbital part of the pairing function satisfying $\langle |{\hat \Delta}_p|^2 \rangle_{\hat p} = 1$. 
In the expression of $\mu$, the small imaginary part $w_z$ ($= O(J/E_F) w_y$) was introduced as a 
cutoff to avoid a possible failure of the perturbation in $J$, although we find that, except in the close vicinity of $T=0$, teh presence of such a small imaginary term does not lead to any quantitatively visible contribution. Hereafter, $\mu$ will be replaced by $w_y/(\sqrt{2} T_c r_H)$. 

Next, $\Delta_n(y,z)$ will be expressed in terms of the Abrikosov lattice solution $\Psi_{\rm A}^{(n)}$, generalized to the $n$-th LL and commensurate to the layer structure, in the form $\alpha^{(n)} \Psi_{\rm A}^{(n)}$, where 
\begin{eqnarray}
\Psi_{\rm A}^{(0)} = \biggl(\frac{\gamma}{\pi}\biggr)^{1/4} (k r_H)^{1/2} \sum_m \exp\biggl({\rm i}kmy - \frac{\gamma}{2 r_H^2} (z - k r_H^2 m)^2 + {\rm i}\frac{\pi}{2}m^2 \biggr),
\label{absn1}
\end{eqnarray}
where $k r_H^2 = k/(2e H) = w d$. 
Then, ${\cal F}_{\rm LD}^{(n)}$ can be rewritten in the form  
\begin{eqnarray}
{\cal F}_{\rm LD}^{(n)} = \Lambda_{2,n} |\alpha^{(n)}|^2 + \frac{V_{4,n} \, \beta_{A,n}}{2} |\alpha^{(n)}|^4,
\label{localLD}
\end{eqnarray}
where 
\begin{eqnarray}
\Lambda_{2,0} &=&  a_0 + 2 c_0 \biggl[ 1 - (N_0(p))^{-1} e^{-p/4} \biggl( \sum_m {\rm cos}(\pi m) \exp\biggl(-\frac{\pi^2 m^2}{p} \biggr) \biggr) \biggr], \nonumber \\
\Lambda_{2,1} &=&  a_1 + 2 c_1 \biggl[ 1 - (N_1(p))^{-1} e^{-p/4} \biggl( \sum_m {\rm cos}(\pi m) \biggl(1 - \frac{p}{2} - \frac{2 \pi^2 m^2}{p} \biggr) \exp\biggl(-\frac{\pi^2 m^2}{p} \biggr) \biggr)\biggr], 
\end{eqnarray}
\begin{eqnarray}
N_0(p) &=& \sum_m \exp\biggl(-\frac{\pi^2}{p}m^2 \biggr), \nonumber \\ 
N_1(p) &=& \sum_m \biggl( 1 - \frac{2 \pi^2}{p}m^2 \biggr) \exp\biggl(-\frac{\pi^2}{p}m^2 \biggr),
\end{eqnarray}
and $p= \pi^2 d^2/r_H^2 = 2 \pi^2 e d^2 H$. 
The positive constant $\beta_{A,n}$ will be defined later. 

If the $H_{c2}$-transition is of second order, the $H_{c2}(T)$-curve consists of a sequence of those satisfying $\Lambda_{2,n}=0$ at the {\it highest} field. However, one should note that, if $V_{4,n} < 0$ on such a line determined by $\Lambda_{2,n}=0$, the $H_{c2}$-transition is discontinuous and should lie just above the line $\Lambda_{2,n}=0$. The coefficient $V_{4,n}$ ($n=0$ or $1$) is given by 
\begin{eqnarray}
V_{4,n} &=& 3 \int_0^\infty d\rho_1 d\rho_2 d\rho_3 \, f\biggl(\sum_{j=1}^3 \rho_j \biggr) \biggl\langle |{\hat \Delta}_p|^4 p_n(\rho_j) \exp\biggl(-\frac{1}{4} \mu^2 \biggl(\sum_j \rho_j \biggr)^2 \biggr) \biggr\rangle, 
\label{v4n}
\end{eqnarray}
where $p_0=1$, and 
\begin{eqnarray}
p_1 = \frac{3}{4}\biggl( 1 - \mu^2 \biggl(\sum_j \rho_j \biggr)^2 + \frac{1}{12} \mu^4 \biggl(\sum_j \rho_j \biggr)^4 \biggr).
\end{eqnarray}
In writing down eqs.(\ref{localLD}) and (\ref{v4n}), wavenumber dependences leading to a spatially {\it nonlocal} interaction between the SC order parameters were neglected \cite{AI}. Such a nonlocality might have brought a subtle change of vortex lattice structure. Hence, this simplification corresponds to assuming that such an effect of nonlocality is much weaker than the effect of the layering on the vortex lattice structure. 

Let us first examine $\alpha_M$-dependences of the mean field $H_{c2}$-line. Typical $H_{c2}$-curves are shown in Figs.1 and 2 (a) by assuming the $H_{c2}$-transition to be of second order. As Fig.1 shows, when $\alpha_M$ is small enough, the $H_{c2}$-curve increases with a positive curvature upon cooling, reflecting the confinement of vortices occurring between the interlayer spacings in higher fields when $\xi_{c0} < d/\sqrt{2}$ \cite{Klemm,II}. The characteristic field $H_{\rm cr}$ beyond which this confinement begins to occur is given by 
\begin{equation}
H_{\rm cr} = \frac{1}{e d^2 \gamma} = 3.6 \gamma H_{\rm 2D}^{({\rm orb})}(0) \biggl(\frac{\xi_c(0)}{d}\biggr)^2,
\end{equation}
where the anisotropy $\gamma = \xi_0/\xi_c(0)$ is conventionally defined in the usual GL region near $T_c$ in terms of eq.(\ref{cn}) in low field limit by 
\begin{equation}
\frac{\xi_c(0)}{d} = \gamma^{-1} \frac{\xi_0}{d} = \frac{\sqrt{7 \zeta(3)}}{8 \pi} \frac{J}{T_c}. 
\end{equation}
In the figures, $h_{cr}=H_{cr}/H^{({\rm orb})}_{\rm 2D}(0) \sim 3.0$. If $H_{c2}(0) > H_{\rm cr}$ due to a large $H_P$ and/or a {\it large} $\gamma$, the saturation of $H_{c2}$ due to the paramagnetic effect at lower temperatures coexists, as in the $\alpha_M=0.01$ case in Fig.1, with the layering-induced positive curvature of $H_{c2}(T)$-line at higher temperatures. In this case, the limiting of superconductivity occurs in the range where the vortices are inactive because they are confined within the interlayer spacings. Then, the approximation in the Pauli limit neglecting the presence of vortices at low temperatures may be useful. However, when $H_{c2}(0) > H_{cr}$, the discontinuous $H_{c2}$ transition and the FFLO state in $n=0$ LL do not easily occur even at low enough temperatures.  Actually, the $H_{c2}$-transition in the $\alpha_M=0.01$ curve remains continuous even in low $T$ limit, and the $n=1$ LL state is never realized 
there. In the intermediate case, $\alpha_M=0.03$, with $H_{c2}(0)$ comparable with $H_{cr}$, the $H_{c2}$-transition becomes discontinuous in $t < 0.29$. Still, the $n=1$ LL instability line (the thick solid curve in Fig.1) lies at lower fields so that the $n=1$ LL state with modulation perpendicular to ${\bf H}$ does not occur. In contrast, if $H_{\rm cr} > H_{c2}(0)$ due to a large $\alpha_M$ and/or a smaller $\gamma$, the limiting of superconductivity occurs in the field range where a slight change of the magnetic field results in structural transitions between different vortex lattices \cite{II,RI02}. In other words, the presence of vortices cannot be neglected in $H_{\rm cr} > H_{c2}(0)$ even within the mean field approximation. As in Fig.2 (a), the $H_{c2}$ curve in this case does not show a portion with a positive curvature at intermediate temperatures. 

\begin{figure}[t]
\scalebox{0.3}[0.3]{\includegraphics{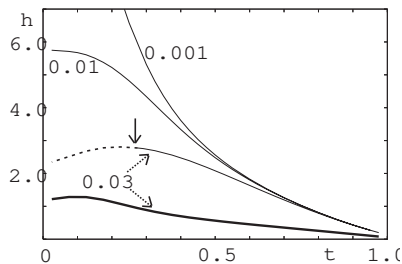}}
\caption{Second order $H_{c2}(T)$-curves (thin solid curves) defined in the $n=0$ LL for $\alpha_M=0.001$, $0.01$, and $0.03$. Here, $h=H/H^{({\rm orb})}_{\rm 2D}(0)$, and $t=T/T_c(0)$. The dotted line is the extrapolation of the thin solid curve for $\alpha_M=0.03$, and the role of the mean field transition $H_{c2}$-transition line is played by a first order transition curve starting from $t=0.275$ indicated by the solid arrow and lying at higher fields. The thick solid curve is the fictitious instability curve in $n=1$ LL for $\alpha_M=0.03$.  } \label{fig.1}
\end{figure}

\begin{figure}[t]
\scalebox{0.3}[0.3]{\includegraphics{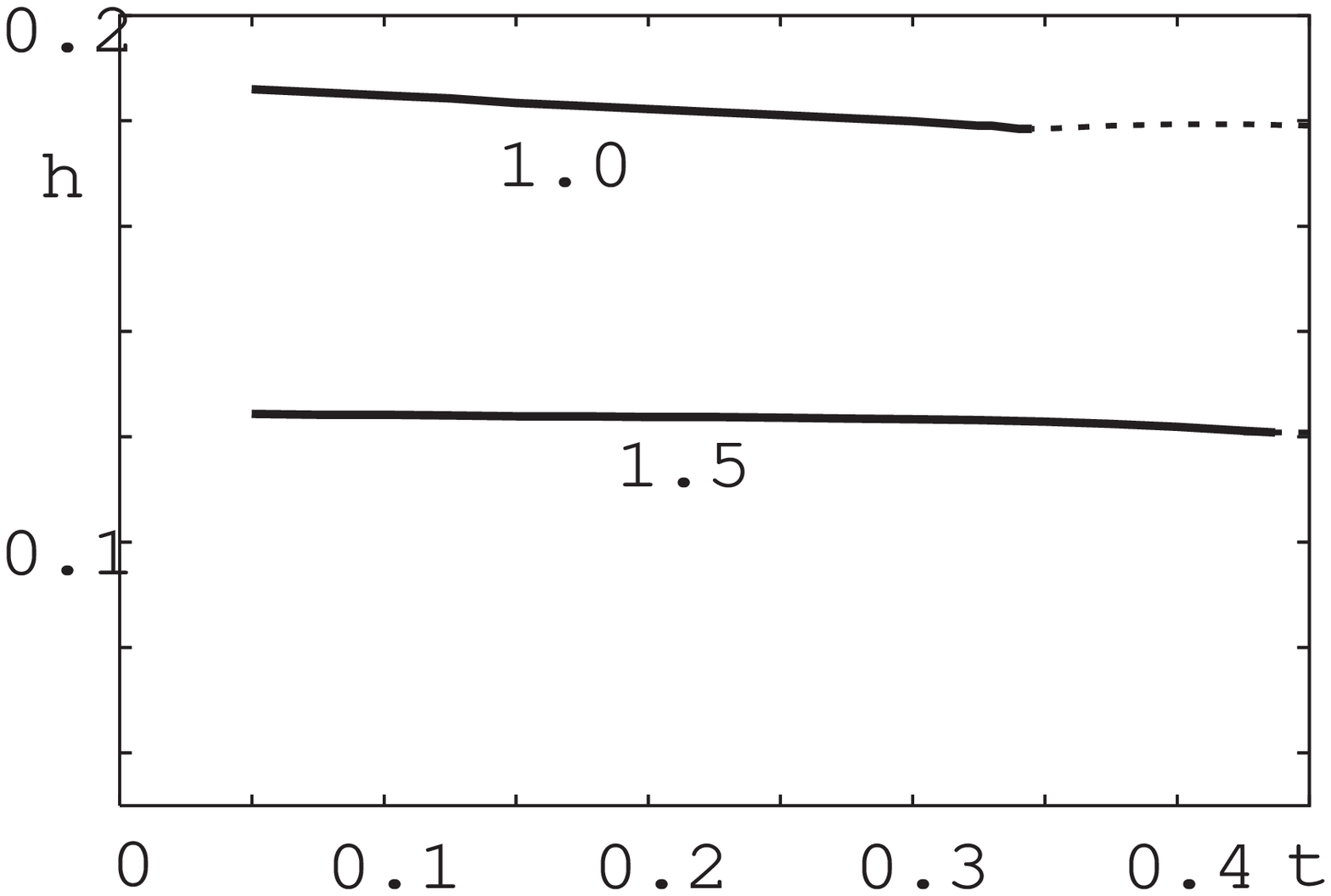}}
\scalebox{0.3}[0.3]{\includegraphics{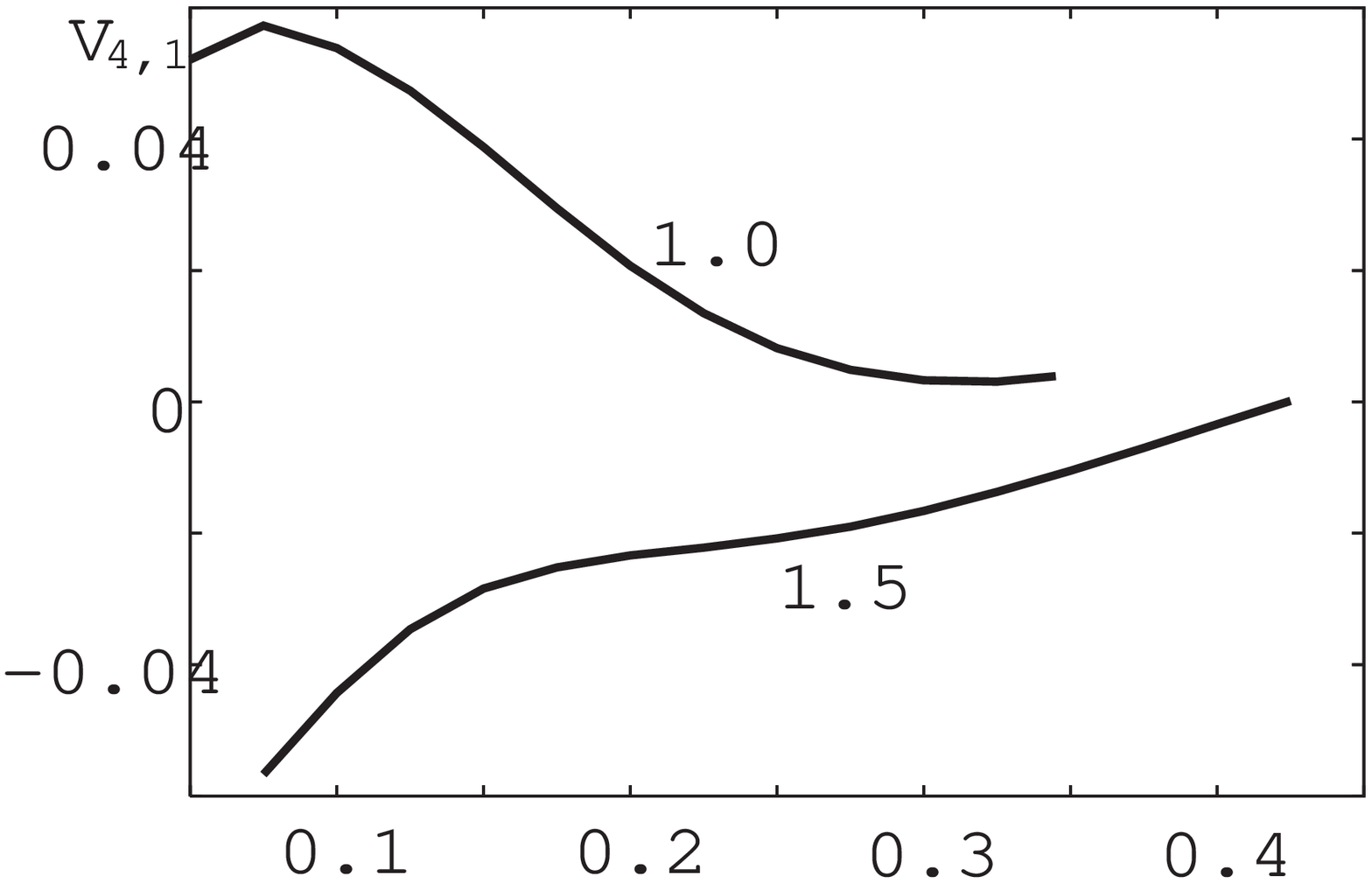}}
\caption{(a) Real and virtual second order $H_{c2}(T)$-transition curves for $\alpha_M=1.0$ and $1.5$. For $\alpha_M=1.0$, the real $H_{c2}$-transition is of second order in $t < 0.3$, and the thin dotted curve is the virtual second order transition curve in $n=0$ LL and is preceded by a first order $H_{c2}$-transition curve lying in slightly higher fields in $0.3 < t < 0.5$. The flat thick solid curve in $\alpha_M=1.5$ is also preceded by a first order $H_{c2}$-transition curve in $n=1$ LL elevating upon cooling. (b) The corresponding $V_{4,1}$ v.s. $t$ curves in $\alpha_M=1.0$ and $1.5$.  } \label{fig.2}
\end{figure}

In contrast, in Fig.2 where $H_{\rm cr} > H_{c2}(0)$, the $n=1$ LL vortex state becomes dominant at lower temperatures : As Fig.2 (a) shows, the $n=1$ LL modes determine $H_{c2}(T)$ and the vortex state just below it in $t < 0.3$ ($t < 0.4$) for $\alpha_M=1.0$ ($\alpha_M=1.5$). Further, the corresponding $V_{4,1}$ curves shown in Fig.2 (b) imply that, for $\alpha_M = 1.0$, the mean field $H_{c2}$ transition is of second order in most of temperatures, while it is rather a discontinuous one for $\alpha_M=1.5$. Just like the $H_{c2}$ line in $n=0$ LL, the $H_{c2}$-transition between the $n=1$ LL state and the normal phase tends to become discontinuous with increasing $\alpha_M$. Since the structural transition between the $n=1$ LL state and any vortex state in $n=0$ LL is inevitably of first order due to the absence of a continuity between their structures, there are two first order or discontinuous transitions in the high field range for the $\alpha_M=1.5$ case. 

Now, let us turn to examining possible vortex lattice structures in the $n=1$ LL state by focusing on the situations, including the $\alpha_M=1.0$ case in Fig.2, with a second order $H_{c2}$ transition. Then, the structure with the lowest value of the positive quartic term of ${\cal F}$ has the lowest energy when a vortex lattice is described in a single LL. Under the assumption neglecting spatial nonlocalities in the quartic term mentioned above, a stable lattice structure of Josephson vortices at each $p$ is determined by the generalized Abrikosov factor 
\begin{equation}
\beta_{A,n} = \frac{\biggl\langle \sum_m \exp\biggl({\rm i}2 \pi m z/d \biggr) |\Delta_1(y,z)|^4 \biggr\rangle_s}{\biggl[ \biggl\langle \sum_m \exp\biggl({\rm i}2 \pi m z/d \biggr) |\Delta_1(y,z)|^2 \biggr\rangle_s \biggr]^2}.
\label{betaa}
\end{equation}
By substituting $\Delta_1 = \alpha_1 \Psi_{\rm A}^{(1)}$ into eq.(\ref{betaa}), it becomes
\begin{eqnarray}
\beta_{A,n} = w \sqrt{\frac{p}{2 \pi}} \sum_{m,s_1,s_2} \frac{B_n(m, s_1, s_2)}{(N_n(p))^2} \exp\biggl(-\frac{\pi^2}{2p}m^2 - \frac{w^2}{2}p(s_1^2+s_2^2) \biggr) \, {\rm cos}(\pi w m(s_1+s_2)+\pi s_1 s_2), 
\end{eqnarray}
where $B_0=1$, 
\begin{eqnarray}
B_1(m,s_1,s_2) &=& \frac{1}{4}(3 - 6 \pi^2 p^{-1} m^2 - 2 w^2 p (s_1^2+s_2^2) + (w^2 p (s_1-s_2)^2 + \pi^2 p^{-1} m^2)(w^2 p (s_1+s_2)^2 + \pi^2 p^{-1} m^2)).
\end{eqnarray}

\vspace{15mm}
\begin{figure}[t]
\caption{Schematic figure expressing the Josephson vortex lattices changing with sweeping the magnetic field $p$. The vortex lattices in the dark regions are accompanied by nodal lines perpendicular to the SC layers (see the figure A in Fig.4), while the nodal planes are parallel to the layers, as in the figure C of Fig.4, in the remaining $p$-ranges. } \label{fig.3}
\end{figure}

A sequence of stable lattice structures following from this $\beta_{A,1}$ is shown in a fixed window of $p$ 
values in Fig.3. In contrast to the Josephson vortex lattices constructed in $n=0$ LL \cite{II,RI02} where diferent lattices are distinguished by the $w$-values, differences in the orientation of nodal planes provide an additional characterization of differerent $n=1$ LL states. The orientation of nodal planes are directly visible in the amplitude of the SC order parameter in $n=1$ LL which is given by
\begin{eqnarray}
|\Delta_1(y,z)|^2 &=& \sum_{m,s} E_{m+s} \biggl(1 - \frac{1}{2}(m^2 (w^2 p) + \pi^2 s^2 (p w^2)^{-1}) \biggr) \, {\rm cos}\biggl(\frac{\pi}{2}m(m+s) \biggr) \nonumber \\ 
&\times& \exp\biggl(-\frac{1}{4} \biggl(m^2 (w^2 p) + \frac{\pi^2 s^2}{p w^2} \biggr) \biggr) \, {\rm cos}(2\pi(m{\overline y} + s{\overline z})), 
\end{eqnarray}
where ${\overline y}=wd \, y/(2 \pi r_H^2)$, and ${\overline z}=z/(2 w d)$, and the sign factor $E_{n}$ is $1.0$ ($0$) for an even (odd) $n$. 
As shown in Fig.4(a), the $\Psi_{\rm A}^{(1)}$ solution at a fixed $w$ ($> 1$) value can become a structure with the lowest energy at two $p$ values. At the low $p$ value, the nodal planes are perpendicular to the layers, while they are oriented along the layers at the higher $p$ value. In general, the nodal planes tend to orient along the direction with a shorter inter-vortex spacing. Since, in the layered system with two-fold anisotropy, the layer structure favors the orientation of vortices parallel to the layers in higher fields, the orientation of nodal planes is affected by the strength of the applied magnetic field. Note that, as Fig.4(b) shows,  the transformation between these two structures at a fixed $w$ does occur not through a rotation of nodal planes but via some merger between the vortices and the nodal planes. Such an intermediated state, Fig.4(b), composed only of the nodal planes is not 
realized for $w > 1$ due to a structural transition to a state with a different $w$-value. However, in the $w=1$ state where all of the interlayer spacings are occupied by the vortices, it is realized : In this case, the nodal planes do not become parallel to the layers because such nodal planes parallel to the SC layers in $w=1$ case would imply a vanishing of $\alpha_1$ on the SC layers and would lead to a strong energy cost. Consequently, the ground state structure in $p > 0.7$ where only the $w=1$ state is realized is that of Fig.4(b). It is interesting to point out that this structure is a kind of square lattice composed only of the nodal planes and similar to the ground state expected in the model in the vortex free Pauli limit \cite{Combescot}. 

\vspace{20mm}
\begin{figure}[t]
\caption{The $p$-dependence of $\beta_{A,1}$ of the $w=6$ Josephson vortex lattice (see eq.(\ref{absn1})) determining the relative stability of the lattices. The figures (A), (B), and (C) express $|\Delta_1(y,z)|^2$ at three points in the first figure, respectively. } \label{fig.4}
\end{figure}

As shown elsewhere, a misfit from a commensurability condition leads to an energy gain by rotating a symmetry axis of vortices from the layers' orientation. In contrast to the $n=0$ LL case with no nodal planes, however, such a rotated solid \cite{RI02} does not lead to lowering of energy in the presence of the additional nodal planes. This is due partially to the fact that, in higher fields, a pinning of nodal planes due to the layer structure leads to an energy gain. Since this pinning due to the layering overcomes the tendency of rotation induced by a misfit, the energy of each rotated state is almost degenerate with the corresponding nonrotated one. 
This is why we have focused on the nonrotated structures in the figures, although their inclusion does not change our interpretation of experimental observations mentioned below. 

Finally, let us discuss relevance of results given above to recent observations in organic superconductors with strong anisotropy suggestive of the presence of an FFLO state \cite{Uji,Lortz}. As is clear from Fig.2, the expected phase diagram in the present situation with {\it strong} anisotropy is not universal. The assumed strong anisotropy suggests that the vortex tilt modulus in the FFLO state in $n=0$ LL with a modulation parallel to ${\bf H}$ will be significantly reduced so that this state may be fragile \cite{RI2}. Then, the discontinuous nature of of the $H_{c2}$-transition between this FFLO state and the normal phase may be changed into a continuous crossover \cite{AI}. The transition between the Abrikosov and FFLO states in $n=0$ LL, appearing in weakly anisotropic cases, is of second order as far as the quasiparticle's lifetime is long enough \cite{AI,RI1}. However, this FFLO state realized in CeCoIn$_5$ may be preceded by the $n=1$ LL state and thus, may not occur in the case with strong anisotropy. In contrast to this, the transition between the $n=1$ LL state and the $n=0$ LL states is, as mentioned earlier, of first order. As indicated through Fig.2, the character of the $H_{c2}$-transition to the $n=1$ LL modulated state depends on the magnitude of $\alpha_M$ and possibly, also on $\gamma$. In particular, the resulting coexistence of the two discontinuous transitions in the case with large enough $\alpha_M$ seems to be consistent with the recent observation of two transitions accompanied by a hysterisis \cite{com2} and a sharp peak of heat capacity in $\kappa$-(ET)$_2$ Cu(NCS)$_2$ \cite{Lortz}. If so, the high field phase at lower temperatures should have a one-dimensional modulation {\it perpendicular} to the field. On the other hand, it is possible \cite{RI1} that the transition between the Abrikosov lattice and an FFLO state in $n=0$ LL modulating along ${\bf H}$ is of first order in the case with a shorter quasiparticle's mean free path. In this case, the high field phase should have a one-dimensional midulation {\it parallel} to the field. The direction of modulation can be clarified, e.g., through ultrasound measurements of the type performed in Ref.\cite{Watanabe}. 

Another main result in this work is the field-induced change of the orientation of nodal planes in the FFLO state constructed in $n=1$ LL. According to the structural transitions between different vortex lattices illustrated in Fig.3, the nodal planes become perpendicular to the SC layers in some field ranges. In real systems, the nodal planes together with the so-called pancake vortices are trapped by point defects becoming active only on the SC layers as pinning sites. If the nodal plane are parallel to the SC layers, the pinning effect on the nodal planes is negligible even if they sit on the SC alyers on averages. Since a large applied current parallel to the $z$-direction, i.e., perpendicular to the layers, can induce a vortex flow parallel to the layers, the above-mentioned pinning effect should be visible in field dependences of out-of-plane resistivity $R_\perp(T)$ data. 
It is believed that the field dependent oscillatory behavior observed in $R_\perp(T)$ data in the field-induced superconductor $\lambda$-(BETS)$_2$FeCl$_4$ \cite{Uji} is an evidence of this pinning effect of nodal planes induced by structural transitions between different Josephson vortex lattices. This explanation of the phenomena is different from an explanation used in Ref.\cite{Uji} based on the scenario in Ref.\cite{BBM} where structural transitions between the Josephson vortex lattices are assumed to be absent, and the nodal planes are not parallel to the SC layers. At the static level, this corresponds to the $w=1$ vortex lattice, represented by Fig.4(b), in the regime $H_{c2} > H_{cr}$. As mentioned in relation to Fig.1, however, the modulation of the FFLO state in this situation is usually parallel to the applied field. Further, according to the conventional description of a vortex flow based on the time-dependent GL equation \cite{AT},  the vortex flow is nothing but an uniform flow of the SC order parameter itself. Since both the vortices and the nodal planes are parts of the SC order parameter, the assumption of \cite{BBM} a vortex flow under nodal planes at rest contradicts the conventional description of SC dynamics \cite{AT}. In contrast, the present picture is applied to the ordinary situation with a strong paramagnetic depairing in which $H_{c2} < H_{cr}$ and hence, with a flat $H_{c2}$ curve at high temperatures. In the ordinary layered materials under parallel fields, the structural transitions between Josephson vortex lattices are not clearly reflected in resistive data because changes of pinning effects accompanying structural changes of Josephson vortex lattices are small. It seems to us that the significant oscillatory behavior of resistivity in Ref.\cite{Uji} is consistent with a large change of pinning effect due to an orientational change of the extended nodal planes. 

Throughout this paper, we have neglected a possibility of a modulation parallel to ${\bf H}$ within the $n=1$ LL vortex state. This vortex state with a {\it two}-dimensional modulation is expected to occur at lower temperatures in the region domianted by the $n=1$ LL. Further details of possible phase diagrams will be discussed elsewhere. 

\begin{acknowledgements}
\end{acknowledgements}

\end{document}